\documentclass[modern]{aastex63}

\usepackage{xcolor}

\renewcommand{\deg}{^{\circ}}
\usepackage{booktabs}
\received{06/30/2020}
\revised{09/08/2020}
\accepted{09/10/2020}
\submitjournal{ApJ}
\shorttitle{Magnetic flux ropes from PSP Encounters 1 \& 2}
\shortauthors{Chen et al.} 

\graphicspath{{FIGS/}}
\begin{document}

\title{Small-scale Magnetic Flux Ropes in the First two Parker Solar Probe Encounters}

\correspondingauthor{Yu Chen}
\email{yc0020@uah.edu}


\author[0000-0002-0065-7622]{Yu Chen}
\affiliation{Department of Space Science, The University of Alabama in Huntsville, Huntsville, AL 35805, USA}

\author[0000-0002-7570-2301]{Qiang Hu}
\affiliation{Department of Space Science, The University of Alabama in Huntsville, Huntsville, AL 35805, USA}
\affiliation{Center for Space Plasma and Aeronomic Research (CSPAR), The University of Alabama in Huntsville, Huntsville, AL 35805, USA}

\author[0000-0002-4299-0490]{Lingling Zhao}
\affiliation{Center for Space Plasma and Aeronomic Research (CSPAR), The University of Alabama in Huntsville, Huntsville, AL 35805, USA}

\author[0000-0002-7077-930X]{Justin C. Kasper} 
\affiliation{Department of Climate and Space Sciences and Engineering, University of Michigan, Ann Arbor, MI 48109, USA}
\affiliation{Smithsonian Astrophysical Observatory, Cambridge, MA 02138, USA}

\author[0000-0002-1989-3596]{Stuart D. Bale}
\affiliation{Physics Department, University of California, Berkeley, CA 94720-7300, USA}
\affiliation{Space Sciences Laboratory, University of California, Berkeley, CA 94720-7450, USA}

\author[0000-0001-6095-2490]{Kelly E. Korreck} 
\affiliation{Smithsonian Astrophysical Observatory, Cambridge, MA 02138, USA}

\author[0000-0002-3520-4041]{Anthony W. Case} 
\affiliation{Smithsonian Astrophysical Observatory, Cambridge, MA 02138, USA}

\author[0000-0002-7728-0085]{Michael L. Stevens} 
\affiliation{Smithsonian Astrophysical Observatory, Cambridge, MA 02138, USA}

\author{John W. Bonnell}
\affiliation{Space Sciences Laboratory, University of California, Berkeley, CA 94720-7450, USA}

\author[0000-0003-0420-3633]{Keith Goetz}
\affiliation{School of Physics and Astronomy, University of Minnesota, Minneapolis, MN 55455, USA}

\author{Peter R. Harvey}
\affiliation{Space Sciences Laboratory, University of California, Berkeley, CA 94720-7450, USA}

\author[0000-0001-6038-1923]{Kristopher G. Klein}
\affiliation{Lunar and Planetary Laboratory, University of Arizona, Tucson, AZ 85721, USA}
\affiliation{Department of Planetary Sciences, University of Arizona, Tucson, AZ 85719, USA}

\author[0000-0001-5030-6030]{Davin E. Larson}
\affiliation{Space Sciences Laboratory, University of California, Berkeley, CA 94720-7450, USA}
 
\author[0000-0002-0396-0547]{Roberto Livi}
\affiliation{Space Sciences Laboratory, University of California, Berkeley, CA 94720-7450, USA}

\author[0000-0003-3112-4201]{Robert J. MacDowall}
\affiliation{Solar System Exploration Division, NASA Goddard Space Flight Center, Greenbelt, MD 20771, USA}

\author[0000-0003-1191-1558]{David M. Malaspina}
\affiliation{Laboratory for Atmospheric and Space Physics, University of Colorado, Boulder, CO 80303, USA}

\author[0000-0002-1573-7457]{Marc Pulupa}
\affiliation{Space Sciences Laboratory, University of California, Berkeley, CA 94720-7450, USA}

\author[0000-0002-7287-5098]{Phyllis L. Whittlesey}
\affiliation{Space Sciences Laboratory, University of California, Berkeley, CA 94720-7450, USA}


%
%
%
%
%
%
%



\begin{abstract}
Small-scale magnetic flux ropes (SFRs) are a type of structures in the solar wind that possess helical magnetic field lines. In a recent report \citep{Chen2020}, we presented the radial variations of the properties of SFR from 0.29 to 8 au using in situ measurements from the Helios, ACE/Wind, Ulysses, and Voyager spacecraft. With the launch of the Parker Solar Probe (PSP), we extend our previous investigation further into the inner heliosphere. We apply a Grad-Shafranov-based algorithm to identify SFRs during the first two PSP encounters. We find that the number of SFRs detected near the Sun is much less than that at larger radial distances, where magnetohydrodynamic (MHD) turbulence may act as the local source to produce these structures. The prevalence of Alfv\'enic structures significantly suppresses the detection of SFRs at closer distances. We compare the SFR event list with other event identification methods, yielding a dozen well-matched events. The cross-section maps of two selected events confirm the cylindrical magnetic flux rope configuration. The power-law relation between the SFR magnetic field and heliocentric distances seems to hold down to 0.16 au.
\end{abstract} 



\section{Introduction} \label{sec:intro}
The magnetic field exists everywhere in the solar wind. The magnetic field or field components from time-series data sometimes can twist, rotate, and form helical lines. This type of structure is called the magnetic flux rope. Based on the scale sizes, it can be categorized into two groups: large-scale flux ropes, i.e., magnetic clouds, and small-scale ones, although the size distribution of flux ropes in the solar wind is believed to be continuous. In contrast to the magnetic cloud, which has an unambiguous solar origin corresponding to coronal mass ejection (CME), and possesses well-defined observational signatures, the source and characteristics of small-scale flux rope (hereafter, SFR) are still under investigation.  

SFRs have been shown via many studies \citep[see, e.g.,][]{Feng2007,Cartwright2010,Zank2014,leroux2015a,leroux2015b,Zhao2018, Hu2018} to correlate with particle energization and some other solar wind structures, such as the interplanetary shock waves, the heliospheric current sheet (HCS), and the stream interaction regions. They are often believed to be associated with magnetic reconnection as well. However, the main question regarding this structure, i.e., where it originates, is still inconclusive. Earlier statistical analysis via various spacecraft measurements but based on limited sample sizes suggested that the SFRs may be generated by small CMEs, and magnetic reconnection across HCS, in both solar corona and interplanetary medium \citep{Feng2008, Cartwright2008, Yu2014}. Moreover, the quantitative analysis of tens of thousands of identified flux tubes via the ACE spacecraft measurements also reveals the classical view that these structures form a packed ``spaghetti''-like configuration owing to processes on the Sun \citep{Borovsky2008,Bruno2001}.

More recently, two-dimensional (2D) magnetohydrodynamic (MHD) turbulence is also considered as the possible source to generate these SFRs. \cite{Greco2009a,Greco2009b,Pecora2019} proved that the current sheets, acting as ``walls'' of SFRs, naturally form during the dynamic evolution process of the solar wind turbulence, correspond well to boundaries of SFRs, by both simulation and observational studies. Furthermore, \cite{Zank2017} concluded that SFRs or vortex structures are a nonlinear component of 2D MHD fluctuations. This view is supported by the observational analysis of SFRs at 1 au. Using the Wind spacecraft measurements, \cite{zhengandhu2018, Hu2018} provided substantial evidence in that there exists ubiquitous SFRs and they correspond to the inertial range of solar wind turbulence.

Such considerable amount of observational analysis is carried out by the Grad-Shafranov-based (in short, GS-based) computer program. This GS reconstruction technique was first developed by \cite{Sonnerup1996} and then first applied to flux rope structures in the solar wind by \cite{Hu2001,Hu2002}. It can determine the flux rope orientation and recover the 2D cross-section from single spacecraft data. Later, the core procedures in this technique were adapted to identify SFRs automatically using time-series data \citep{zhengandhu2018,Hu2018}. This program succeeded in detecting 74,241 SFRs from 21-year worth of Wind spacecraft measurements \citep{Hu2018}. From this immense number of events, those studies provided strong observational evidence for a scenario that SFR being generated locally from MHD turbulence. Lately, it was extended to the full Helios, ACE, Ulysses, and part of the Voyager spacecraft measurements and four individual SFR databases were produced covering the heliocentric distances from 0.29 to $\sim$7-8 au on both the ecliptic plane and over mid- to high-latitude regions \citep{Chen2019,Chen2020}. These databases assisted in the investigation of SFR properties at different heliographic latitudes and heliocentric distances. 

The Parker Solar Probe (PSP) mission, launched on 2018 August 12, is a man-made spacecraft that shatters the record of approaching the closest heliocentric distance to the Sun. It is designed to have a total of 24 orbits around the Sun in the ecliptic, and the final orbit will approach a distance about 0.046 au to the Sun. With these close encounters near perihelion, it aims to investigate the low solar corona, the source region of solar wind, and the dynamics leading to the formation of supersonic and super-Alfv\'enic solar wind streams \citep{Fox2016}. In the first two encounters, the PSP passed the perihelion on 2018 November 5 and 2019 April 4, respectively, at a distance $\sim$0.16 au from the Sun. As aforementioned, the previous closest encounter was achieved by the Helios spacecraft mission at a radial distance $\sim$ 0.29 au decades ago. With the launch of the PSP spacecraft and the range of heliocentric distances it will cover, the investigation of SFR characteristics in the inner heliosphere (with the heliocentric distances less than 0.29 au) becomes feasible, starting with the first two encounters. 

On one hand, this new observational point may be able to provide more information about the source of SFRs. In light of a series of GS-based detection results, SFRs tend to be generated locally in the MHD turbulence at 1 au and farther distances, where the direct solar source effect may not be strong enough. On the contrary, the chance for a spacecraft passing through the solar originated SFRs is greatly higher at smaller heliocentric distances via, e.g., the PSP measurements at closer encounters. SFRs with separate origins probably have different inherent signatures. Whether the consistency between SFRs at $\geqslant$ 1 au and near the Sun still exists can assist us to examine further on the aforementioned question on the inconclusive source of SFRs. On the other hand, the radial variations of SFR properties can also be extended further to the inner heliosphere. According to our experiences, the existence of SFR can be confirmed by various observational techniques, such as the wavelet analysis \citep{Zhao2020}, the minimum variance analysis of magnetic field (MVAB; \cite{Sonnerup1998}) (e.g., \cite{Yu2014}). SFR boundaries, however, may vary among different techniques. Therefore, in this study, we continue to apply the GS-based program to PSP measurements in order to guarantee self-consistent SFR detection results and enable comparisons with existing SFR databases obtained by the same approach. Such comparisons will also shed light on the question regarding the origin by examining whether any trend may extend to smaller radial distances in a persistent manner. 

This paper is organized as follows. The GS-based automated program and the core characteristics for this SFR detection will be introduced briefly in Section \ref{sec:method}. Also, the detection period, criteria, and PSP data processing will be presented. The main detection results are presented in Section~\ref{sec:results}. In Section \ref{sec:overview}, an overview of the basic parameters, such as the magnetic field, the Alfv\'en speed, the plasma properties, etc., together with identified SFR structures are shown for the full detection periods of encounter 1 and part of encounter 2. In a series of recent papers, additional (generally large-scale) structures, such as the HCS, magnetic reconnection signatures, and an ICME, etc., were reported and we discuss their associations with the occurrence of some SFRs in Section \ref{sec:cor}. The result of this paper is also compared with others which use totally different methods. In Section \ref{sec:radial}, earlier detection result in \cite{Chen2020} is cited and combined with the result via PSP dataset. The comparison between PSP and Helios SFR lists, albeit limited, is presented, and the radial variations are shown. Finally, the main findings and future work are summarized in the last section. 

\section{Method and Data} \label{sec:method}

The method for detection of SFRs in this study is a GS-based automated computer program. In this program, the main feature we are seeking for an SFR in time series data array, based on the GS reconstruction technique, is the double-folding pattern in the relation between two physical quantities, namely, the transverse pressure $P_t$ and the magnetic flux function $A$ (also the axial component of the magnetic vector potential). All are calculated from in-situ spacecraft measurements. The transverse pressure $P_t$ is the sum of the thermal pressure $p$ and the axial magnetic pressure $B_z^2/(2\mu_0)$, and the magnetic flux function $A$ can be acquired by integrating the 1D magnetic field component \citep{Hu2017GSreview}. The standard GS equation prescribes that the flux function $A$ acts as the single variable of $P_t$. This one-to-one correspondence allows the determination of a 2D flux rope configuration characterized by a set of nested isosurfaces of $A$. On each isosurface, the corresponding values of $P_t$ remains the same. Therefore the search of such a configuration is facilitated by examining the $P_t$ versus $A$ arrays for the double-folding and single-valued pattern of $P_t(A)$. The goodness of the satisfaction of the single-value function relation $P_t(A)$ is judged by a set of quantitative criteria including a fitting residue as a result of an analytic fitting function $P_t(A)$ to the data. For examples of such relations, see Section~\ref{sec:cor}. 

During the process of data scanning and calculations of relevant quantities, any array owns the double-folding pattern will be saved as a potential candidate. Notice that a flux rope candidate would not only need to be double-folded in $P_t(A)$ arrays but also have a good quality of folding/overlapping. This requires that the data points split into two branches which appear to fold back with one branch approximately overlapping on top of the other. Therefore, the two residues, which evaluate the difference between the two folding branches as well as a fitting residue of the $P_t(A)$ function, are adopted to ensure the quality of overlapping \citep{Hu2002, Hu2004}. Last but not least, the low Alfv\'enicity, i.e., the correlation between fluctuations of the magnetic field and velocity \citep{Belcher1971}, is required to distinguish flux ropes from other highly Alfv\'enic structures, such as torsional Alfv\'en waves. This is implemented through a threshold condition on the Wal\'en test slope, which is derived from the linear regression between the remaining flow velocity and the local Alfv\'en velocity in a component-wise way \citep{Paschmann1998}.

We examine PSP data for the first two encounters from 2018 October 31 to December 19 and from March 7 to May 15, 2019, respectively. The magnetic field and plasma data are measured by two instrument suites onboard: the FIELDS Experiment \citep{Bale2016} and the Solar Wind Electrons Alphas and Protons (SWEAP) \citep{Kasper2016,Case2020,Whittlesey2020}. All data used in this study are those tagged by ``Only Good Quality'' on the NASA CDAWeb. Due to different cadences, the magnetic field data and plasma data are not always in accordance with each other in time-series. In order to bridge this inconsistency, a down-sampling process (averaged to the lower sample rate) is applied to the magnetic field data (sometimes also to the plasma data) to match these two datasets and keep the original plasma data as much as possible. The combined magnetic field and plasma dataset for analysis has a uniform cadence of 28 seconds.

\begin{table}
\begin{center}
\caption{Criteria of SFR detection via PSP dataset.}
\begin{tabular}{lcc}
\toprule
PSP & Encounter 1 & Encounter 2\\
\midrule
Time Period & Oct 31 - Dec 19, 2018 & March 7 - May 15, 2019\\
Duration Range (min) & \multicolumn{2}{c}{5.6 $\sim$ 360} \\
Wal\'en Test Slope Threshold & \multicolumn{2}{c}{0.3} \\
$\langle B\rangle$ (nT) & $\geqslant$ 25 & $\geqslant$ 10\\
\bottomrule
\end{tabular}
\label{table:criteria}
\end{center}
\end{table}

Table \ref{table:criteria} lists the criteria of SFR detection for the first two encounters of the PSP. The detection is carried out for about one and a half months for each round. The SFR duration range is set from about 6 to 360 min. For a flux rope candidate, the duration is the time interval length for a spacecraft crossing the structure. From the aspect of the detection algorithm, it represents the lower limit of the data segment length of the double-folding parts within a corresponding searching window. In other words, we assume that the spacecraft should have spent at least 6 min to cross the shortest flux rope and about 360 min (6 hr) for the longest one. In this study, all possible small-scale flux rope candidates are assumed to be located within this range.

According to the recent reports, an abundance of Alfv\'enic signatures or Alfv\'en wave-like structures in the predominantly slow wind were observed within fluctuations at distances closer to the Sun during the first two PSP encounters. The duration of these structures is up to several minutes \citep{Kasper2019,Bale2019}. They may exhibit similar magnetic field components rotation as a flux rope, but often with the significant field-aligned flow, which can be characterized by the Wal\'en test in order to be distinguished from a flux rope in quasi-static equilibrium. The exclusion of similar structures other than flux ropes is crucial. In our study, the Wal\'en test slope is employed to discern whether a structure has high Alfv\'enicity. According to its definition, i.e., the ratio of the remaining flow speed to the local Alfv\'en speed, we set 0.3 as a lower threshold based on our experiences to diminish the effect of the Alfv\'enic structures. In addition, this threshold value 0.3 was used in the prior studies to indicate the level of the significant remaining flows \citep{Hu2018}. It was also used to establish the condition for the magnetohydrostatic equilibrium \citep{Hasegawa2014} from which the GS equation is derived.

Additionally, although the SFR is small scale in nature relative to the magnetic cloud, it is large enough when compared to magnetic fluctuations in the background. Consequently, the removal of these noises is also essential. Considering that the range of radial distances for encounter 2 in this detection is wider than encounter 1, the lower limits of the average magnetic field magnitude based on the Parker magnetic field are set, respectively, i.e., as 25 and 10 nT.

\begin{table}
\begin{center}
\caption{Detection result of SFRs via PSP dataset.}
\begin{tabular}{lcc}
\toprule
SFR Occurrence & Oct 31 - Nov 15, 2018 & March 8 - Apr 18, 2019\\
\midrule
Total Count & 24 & 20 \\
Radial Distance (au) & 0.1717 $\sim$ 0.3199 & 0.1662 $\sim$ 0.6615\\
Scale Size (au) & 0.0004 $\sim$ 0.0215 & 0.0003 $\sim$ 0.0316\\
Duration (min) & 5.6 $\sim$ 165.6 & 5.6 $\sim$ 276.3\\
\bottomrule
\end{tabular}
\label{table:result}
\end{center}
\end{table}

\section{SFR Detection Results via PSP Dataset}\label{sec:results}

\subsection{Overview}\label{sec:overview}

Table \ref{table:result} summarizes selected SFR parameters from the detection results for the two encounters, including the total count numbers, and the ranges of radial distances, scale sizes, and duration. The total numbers of SFRs detected in the first two encounters are 24 and 20 respectively. In these records, SFRs were found to locate within a wide range of heliocentric distances from 0.16 to 0.66 au, and have a distinct scale size range. The smallest SFR (in scale size) is 0.0003 au for both encounters, and the longest SFR (in duration) is 276.3 min ($\sim$ 4.6 hr).


\begin{figure}
\centering
\includegraphics[width=1.0\textwidth]{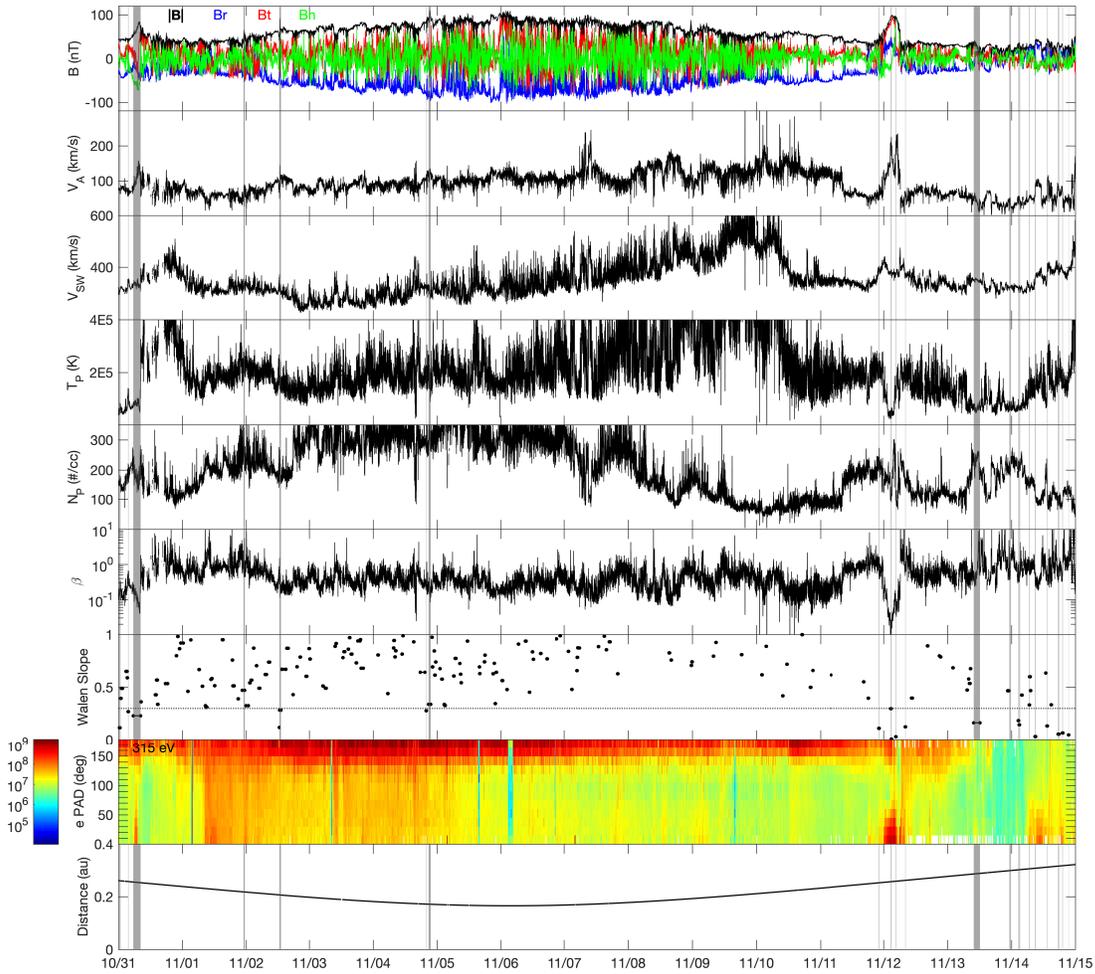}
\caption{Time-series plot from 2018 October 31 to November 15. From the top to the bottom panels are the magnetic field magnitude and components in the $RTN$ coordinates, the Alfv\'en speed, the solar wind speed, the proton temperature, the proton number density, the proton $\beta$, the Wal\'en test slope for each SFR candidate, the electron pitch angle distribution (PAD) for the 315 eV energy channel, and the PSP radial distance in au. In the 7th panel, the Wal\'en slope threshold 0.3 is denoted by the horizontal dot-dashed line. Across all panels, the identified SFR intervals are marked by gray shaded areas.}\label{fig:overview1}
\end{figure}

During the detection in PSP encounter 1 from 2018 October 31 to 2018 December 19, there are 24 SFRs identified by the GS-based program totally. These SFRs concentrate within a period of half a month. Figure \ref{fig:overview1} presents the time-series data from 2018 October 31 to November 15, {which includes the magnetic field components in the $RTN$ (radial, tangential, and normal) coordinates, the plasma parameters (the proton temperature $T_p$, the proton number density $N_p$, and the proton $\beta$, etc.), the Wal\'en test slope, the electron pitch angle distribution (PAD), and the radial distance of the spacecraft. The Wal\'en test slope is given for each flux rope candidate before applying the threshold condition. The final identified quasi-static SFRs are marked by gray shaded areas across all panels.} Most shaded areas may appear like vertical lines because of their relatively short duration compared with the range of the horizontal axis over the period of fifteen days.

Note that these identified SFR intervals marked in Figure~\ref{fig:overview1} are obtained by applying the Wal\'en test slope threshold 0.3. They seem to distribute unevenly across the time period during encounter 1. There are clearly more SFR candidates but with higher Wal\'en test slope, $>0.3$, as indicated in the 7th panel.

The magnetic field has an apparent increase in strength when the spacecraft probed gradually closer to the perihelion on 2018 November 6. Few SFRs are detected around this time. When PSP traveled away from the perihelion, starting from late November 11, SFRs occur more frequently while the number of SFR candidates begin to decrease (7th panel, Figure~\ref{fig:overview1}). In general, the plasma properties, such as $T_p$, $N_p$, and plasma $\beta$, do not have consistently coincident variations with the corresponding SFR intervals. This phenomenon was also observed by \cite{Yu2014}, for instance, who found that highly suppressed $T_p$ and low plasma $\beta$ do not represent the typical characteristics for SFRs. It is also seen, as a general trend, that the identified SFRs tend to occur in rather slow solar wind streams, with $V_{SW}\approx 300$ km/s. During a brief time period past perihelion when the solar wind speed exceeded 400 km/s, the event candidates are much fewer. 

The eighth panel in Figure \ref{fig:overview1} shows the electron pitch angle distribution. Following \cite{Nieves2020}, the energy channel 315 eV is selected as it may indicate the electron streaming direction with respect to the local magnetic field connecting back to the Sun presumably. The bidirectional enhancement of the electron PAD at both 0$\deg$ and 180$\deg$ is not very common for most of the SFR records in this encounter. Such signatures of enhanced bidirectional electron PAD are most prominent for the ICME event and the HCS crossing that occurred on 2018 November 12 and 13, as discussed in \citet{Giacalone2020} and \citet{Szabo2020}, respectively.

\begin{figure}
\centering
\includegraphics[width=1.0\textwidth]{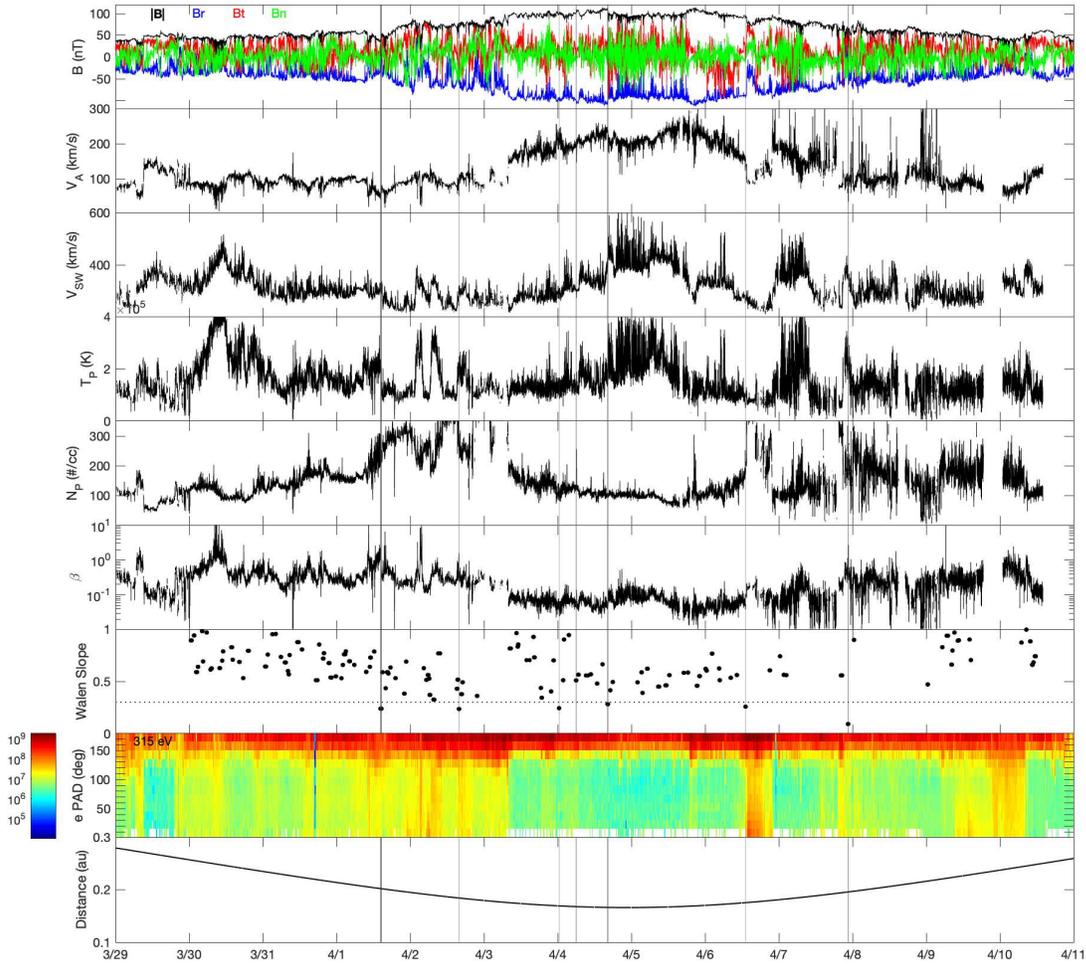}
\caption{Time-series plot from 2019 March 29 to April 11. The format follows that of Figure \ref{fig:overview1}.}\label{fig:overview2}
\end{figure}

The detection for the PSP encounter 2 is implemented for $\sim$ 2 months from 2019 March 7 to 2019 May 15. During this period, the total number of identified SFRs is 20, which is a little less than that from the first encounter. Notice that the automated detection demands complete and continuous data coverage as much as possible. The interpolation by using the surrounding values is the common approach for handling small data gaps. Nonetheless, it is not appropriate for large and successive gaps, especially for the main parameters in this detection, e.g., the solar wind velocity, which is essential for obtaining the flux rope frame velocity. Unfortunately, the data gaps may occur more frequently for in-situ measurements with ultra-high cadence, such as the case for PSP. This data issue may lead to the abnormal result that fewer SFRs are identified during the longer detection period instead of more events.

Figure \ref{fig:overview2} presents the same PSP measurements and detection results around the perihelion during encounter 2. Compared with results in encounter 1, the magnetic field also has increased strength when the PSP was closer to the perihelion, but with relatively less rapid fluctuations in the $B_t$ and $B_n$ components. The plasma $\beta$ is also much depressed around the perihelion. The other different aspect from encounter 1 result is the occurrence of SFR candidates around the perihelion. Albeit the number of structures with high Alfv\'enicity is still far more than that of SFRs, the Wal\'en test slope values are generally smaller. There are three SFR events identified around the perihelion. Moreover, the variability in the electron PAD also appears to be greater than in encounter 1. There are markedly some SFRs in encounter 2 associated with the electron PAD enhancement nearby. The topic of the correspondence between SFRs and the associated electron PAD signatures remains an important one and has yet to be further studied in the era of the PSP and the Solar Orbiter missions. The generation of the list of SFR events as we strive to do here will assist in this specific study and other relevant studies.

\subsection{Occurrence Rate }\label{sec:occur}
By adopting the GS-based technique, a total number of 44 SFRs were identified during the first two PSP encounters. Although the implemented detection is aimed for 1.5 months at least per orbit, the distributions of SFRs cluster within a half to one month due to incomplete data coverage.
In \cite{zhengandhu2018}, the SFR database via the Wind spacecraft measurements from 1996 to 2016 (covering two solar cycles) yields an average monthly count of SFRs about 294 at 1 au. By combining the results of two encounters and considering the total time periods, the equivalent monthly count via PSP is still notably fewer, i.e., about 27 per month. As shown in the previous subsection, this is a result of the strict Wal\'en test slope threshold we implemented. 

This discrepancy in counts due to enhanced Alfv\'enicity closer to the Sun may be ascribed to the possible solar source of SFRs. One traditional view suggests that SFRs originate as streamer blobs of variable sizes that can be traced back to the Sun. This type of structures is usually observed by white-light coronagraphic imaging, and has been reported by such remote-sensing observations via PSP and STEREO \citep{Korreck2020,Nieves2020}. 

An alternative view infers that these small structures could be generated locally, which is correlated to a cascade of 2D MHD turbulence \citep{zhengandhu2018,Hu2018,Pecora2019}. To some extent, this view can reconcile the contradiction of the number of SFRs at different locations in the heliosphere. A group of SFRs originating from the Sun can be recognized more easily at places closer to the Sun, while the turbulence source for generating a part of SFRs may be dominating at farther distances as it was suggested that turbulence may become more important at farther radial distances \citep{Matthaeus1990}. Although the Sun may still have noticeable consequences on the SFR occurrence and properties at these distances, this impact may not be as powerful as the effect due to the solar wind turbulence and local dynamic processes \citep[e.g.,][]{Zank2014}. For example, the solar cycle dependency of SFR occurrence is obvious at 1 au \citep[see, e.g,][]{zhengandhu2018}, but such dependency is modulated by variations with both the radial distances and the heliographic latitudes at farther distances \citep{Chen2019}.

In \cite{Chen2020}, the radial variation of SFRs was investigated by examining one-year detection results via Helios, ACE, Ulysses, and Voyager datasets from 0.29 to about 7-8 au at low latitude regions. Those authors reported that SFR count decays with increasing radial distance $r$ due to the possible merging process, which follows a power-law with an index -0.77 (or a deduced index around -1.5 for a 2D region). In the situation of solar origin, one would speculate the occurrence rate of SFRs to reduce with increasing $r$ in a consistent way. Such a trend if any has yet to be further quantified with more SFR events from PSP, especially for $r<0.29$ au.

Despite the uncertainty regarding the source of SFRs, strict criteria of detection can result in the suppression of the identification of SFRs. At the present time, one of these important limits is to remove the flux-rope-like structure, i.e., highly Alfv\'enic ones. \cite{McComas2000} found that the Alfv\'enic structures are more likely to prevail in the fast wind (including low latitudes) and at high latitudes. The detection of SFRs via Ulysses measurements also unveils that flux ropes at high latitude regions are accompanied more often by those structures \citep{Chen2019}. Despite the common belief that higher Alfv\'enicity occurs more likely in the fast wind than in the slow wind, PSP measurements during the first two encounters have manifested lots of Alfv\'enic turbulence close to the perihelion while traveling mostly in the slow wind. 

As introduced in the previous section, the Wal\'en test slope threshold is applied as the condition in this analysis to exclude the Alfv\'enic structures. A value 0.3 is also adopted for 1 au detection. For example, the total number of SFRs in the year 2004 via ACE dataset is 3049. This number becomes 3162 when relaxing this value to be 1.0, i.e., including all structures that possessing the signature of double-folding $P_t(A)$, no matter how high the Alfv\'enicity of the structure is. It could be expected to have such a small reduction (3\%) in the total count at 1 au since the Alfv\'enicity may reduce outward to farther distances and could only survive in the fast wind \citep[e.g.,][]{Panasenco2020}. 
However, the threshold value 0.3 becomes critical for the PSP detection when identifying SFRs with the same procedures but under the circumstances with lots of highly Alfv\'enic structures. The total number of SFRs would be 175 and 209, respectively, during these two encounters when loosening this limit to 1.0. Now, the equivalent monthly count becomes 230, which is comparable to 1 au detection result. In other words, 88\% of possible SFR candidates are eliminated due to high Alfv\'enicity during the first two encounters. This raises a question regarding the essence of these Alfv\'enic structures. For instance, the torsional Alfv\'en wave embedded within a small flux rope was reported by \citet{Gosling2010} at 1 au, but it is extremely rare. The ongoing PSP observations offer the opportunity to look for and further characterize these structures.

\subsection{SFR Occurrence in Conjunction with Other Structures}\label{sec:cor}

With the release of PSP data in the first encounter, a variety of solar wind structures, such as the HCS, magnetic reconnection regions, and an ICME, etc., were reported in a number of recent papers. The correlation and validation, therefore, can be examined by comparing the event list in this paper with other studies. Table \ref{table:full} lists all identified SFR events with the time interval, duration, scale size, the average magnitude of the SFR magnetic field, the sign of the magnetic helicity ($\sigma_m$), and the axis orientation. Here, the magnetic helicity of each flux rope is consistent with one particular parameter which is derived within each interval, i.e., the product of the magnetic flux function $A$ and the axial field component $B_z$, as a proxy to the magnetic helicity density. The sign of the extremum is either positive or negative in the array of $A \cdot B_z$, which yields the sign of magnetic helicity of the corresponding SFR. The axis orientation is given by two directional angles (in degrees), the polar angle $\theta$ and the azimuthal angle $\phi$. They describe the angle between the flux rope cylindrical axis $\hat\mathbf{z}$ and the local $N$-direction, and the angle between the $R$-direction and the projection of $z$-axis onto the $RT$-plane, respectively. 

\begin{table}[ht]
\small
\caption{List of Small-scale Flux Ropes identified during \emph{PSP} Encounter 1 $\&$ 2.}\label{table:full}
\footnotesize
\centering
\begin{tabular}{lllllcl}
\toprule
\multicolumn{1}{c}{No.} & \multicolumn{1}{c}{Time Interval} & \multicolumn{1}{c}{Duration} & \multicolumn{1}{c}{Scale Size} & \multicolumn{1}{c}{$\langle {B} \rangle$} & \multicolumn{1}{c}{Sign of $\sigma_m$} & \multicolumn{1}{c}{$(\theta, \phi)$}\\
 & \multicolumn{1}{c}{UT} & \multicolumn{1}{c}{(sec)} & \multicolumn{1}{c}{(au)} & \multicolumn{1}{c}{(nT)} & \\
\midrule
1$^*$ & 2018 Oct 31 00:20-00:36 & 981 & 0.0014 & 42.33 & $+$ & (100, 140) \\
2 & 2018 Oct 31 00:49-00:55 & 365 & 0.0006 & 44.01 & $-$ & (60, 140) \\
3$^*$ & 2018 Oct 31 03:36-03:45 & 533 & 0.0011 & 41.00 & $-$ & (120, 120) \\
4$^*$ & 2018 Oct 31 05:33-08:18 & 9941 & 0.0215 & 68.34 & $+$ & (160, 160) \\
5 & 2018 Nov 1 22:56-23:25 & 1737 & 0.0035 & 49.04 & $-$ & (80, 240) \\
6 & 2018 Nov 2 12:23-12:29 & 365 & 0.0008 & 43.28 & $+$ & (20, 200) \\
7 & 2018 Nov 2 12:29-13:21 & 3109 & 0.0064 & 64.86 & $-$ & (30, 140) \\
8 & 2018 Nov 4 19:50-19:57 & 365 & 0.0004 & 86.02 & $+$ & (100, 160) \\
9$^*$ & 2018 Nov 4 20:46-21:34 & 2885 & 0.0064 & 95.89 & $-$ & (130, 260) \\
10 & 2018 Nov 11 22:08-22:19 & 701 & 0.0018 & 51.68 & $-$ & (130, 80) \\
11$^*$ & 2018 Nov 12 02:39-02:47 & 505 & 0.0013 & 95.18 & $+$ & (100, 80) \\
12$^*$ & 2018 Nov 12 02:50-02:58 & 533 & 0.0012 & 97.97 & $+$ & (100, 60) \\
13$^*$ & 2018 Nov 12 04:30-04:40 & 589 & 0.0015 & 94.59 & $+$ & (20, 100) \\
14 & 2018 Nov 12 08:06-08:13 & 449 & 0.0011 & 37.14 & $-$ & (40, 100) \\
15$^*$ & 2018 Nov 13 09:49-12:09 & 8373 & 0.0194 & 27.60 & $+$ & (30, 80) \\
16$^*$ & 2018 Nov 13 23:12-23:41 & 1737 & 0.0031 & 27.66 & $-$ & (50, 320) \\
17$^*$ & 2018 Nov 14 02:32-02:41 & 533 & 0.0011 & 31.21 & $-$ & (20, 40) \\
18$^*$ & 2018 Nov 14 02:53-03:12 & 1121 & 0.0021 & 32.45 & $-$ & (60, 60) \\
19 & 2018 Nov 14 06:37-06:48 & 701 & 0.0005 & 30.81 & $+$ & (100, 20) \\
20 & 2018 Nov 14 08:51-09:04 & 757 & 0.0016 & 41.13 & $-$ & (110, 300) \\
21 & 2018 Nov 14 13:18-13:27 & 533 & 0.0013 & 36.57 & $-$ & (120, 280) \\
22$^*$ & 2018 Nov 14 17:30-17:50 & 1177 & 0.0022 & 34.55 & $-$ & (110, 320) \\
23 & 2018 Nov 14 19:06-19:15 & 561 & 0.001 & 26.22 & $-$ & (110, 320) \\
24 & 2018 Nov 14 21:23-21:28 & 337 & 0.0006 & 30.35 & $+$ & (70, 320) \\
25 & 2019 Mar 7 15:01-15:10 & 533 & 0.0015 & 13.32 & $-$ & (70, 300) \\
26 & 2019 Mar 7 21:59-Mar 8 0:01 & 7365 & 0.0181 & 15.73 &$+$ & (40, 300) \\
27 & 2019 Mar 8 01:16-01:51 & 2101 & 0.0037 & 14.48 & $-$ & (50, 340) \\
28 & 2019 Mar 13 01:19-05:56 & 16577 & 0.0316 & 12.76 & $+$ & (80, 80) \\
29 & 2019 Mar 14 06:00-06:07 & 421 & 0.0008 & 13.95 & $-$ & (30, 220) \\
30 & 2019 Mar 14 17:24-17:30 & 365 & 0.0007 & 12.73 & $+$ & (120, 260) \\
31 & 2019 Mar 15 12:47-13:17 & 1821 & 0.005 & 30.00 & $+$ & (120, 100) \\
32 & 2019 Mar 20 04:12-04:19 & 449 & 0.0006 & 12.72 & $+$ & (130, 220) \\
33 & 2019 Mar 20 15:44-15:50 & 337 & 0.0008 & 13.44 & $+$ & (10, 20) \\
34 & 2019 Mar 27 10:35-10:42 & 421 & 0.0008 & 23.64 & $-$ & (40, 220) \\
35 & 2019 Mar 27 19:21-19:27 & 393 & 0.0008 & 13.71 & $+$ & (20, 220) \\
36 & 2019 Apr 1 14:15-14:39 & 1401 & 0.0026 & 45.24 & $+$ & (80, 120) \\
37 & 2019 Apr 2 15:47-15:56 & 505 & 0.0003 & 68.49 & $-$ & (110, 180) \\
38 & 2019 Apr 4 00:24-00:29 & 337 & 0.0008 & 98.99 & $-$ & (10, 160) \\
39 & 2019 Apr 4 5:55-06:05 & 617 & 0.0012 & 101.98 & $-$ & (130, 140) \\
40 & 2019 Apr 4 16:11-16:25 & 841 & 0.0012 & 107.17 & $+$ & (80, 160) \\
41 & 2019 Apr 6 13:08-13:13 & 337 & 0.0006 & 94.31 & $+$ & (110, 120) \\
42 & 2019 Apr 7 22:29-22:39 & 561 & 0.0009 & 57.89 & $-$ & (110, 320) \\
43 & 2019 Apr 18 13:50-13:56 & 337 & 0.0012 & 14.78 & $+$ & (160, 140) \\
44 & 2019 Apr 18 15:07-15:16 & 533 & 0.0019 & 13.32 & $+$ & (130, 80) \\
\bottomrule
\end{tabular}
\tablenotetext{*}{Indication of the overlapping event with the result in \cite{Zhao2020}.}
\end{table}

In this list, some SFRs occur in association with other specific solar wind structures. For example, events No.7 and No.14 are correlated with magnetic reconnection regions, which were reported in \cite{Phan2020}. Also, 3 SFRs on 2018 November 12 and 13 are probably HCS related events, as demonstrated in \cite{Szabo2020} who discussed the occurrence of HCS traversal by PSP and listed a few possible SFRs with slow rotating magnetic field components. This type of correlation was also found by \cite{Hu2018} via the Wind spacecraft dataset, which showed that the SFRs are more inclined to accumulate near HCS in the slow wind. 

\begin{figure}
\centering
\includegraphics[width=0.45\textwidth]{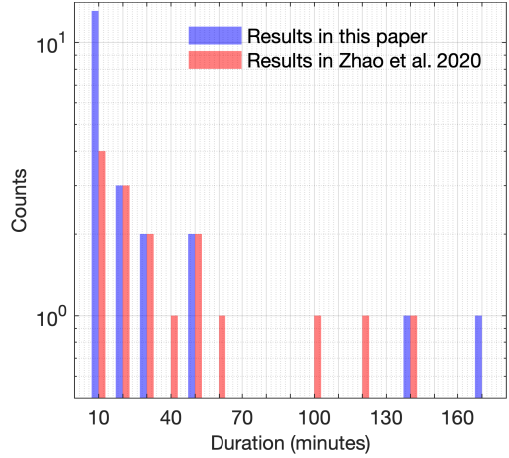}
\caption{Distributions of SFR duration for all events (except for the single ICME flux rope event with the longest duration) in the same detection period from Table~3 and those reported in \cite{Zhao2020}. }\label{fig:compz}
\end{figure}

Alternatively, \cite{Zhao2020} identified 40 SFRs from 2018 October 22 to November 15 by using a wavelet-based analysis. The spectrograms of reduced magnetic helicity, cross-helicity, and residue energy were calculated from the time-series data to support the identification of quasi-static SFR structures. These SFRs are characterized as possessing (all normalized) relatively enhanced magnetic helicity, small cross-helicity, and residue energy close to -1. The latter two conditions correspond to low Alfv\'enicity. Therefore their results comply with our identification of the same type of quasi-static SFRs, governed by the GS equation in our approach. In the same detection period, there are 12 SFRs (event numbers with asterisks in Table \ref{table:full}) in this study coinciding with the list in \cite{Zhao2020}. The signs of magnetic helicity of these overlapping events are the same as those derived by the wavelet analysis. Figure \ref{fig:compz} shows the distributions of SFR duration, which includes all events from these two sets of results in the same detection period. Although the SFR duration is in general a little longer in \cite{Zhao2020} than in this study, smaller duration dominates for both sets, and the temporal scales via the two methods are quite comparable. Among these overlapping events, we select two SFRs as examples and present the time-series data with cross-section maps via the GS reconstruction in what follows. 

\begin{figure}
\centering
\includegraphics[width=0.9\textwidth]{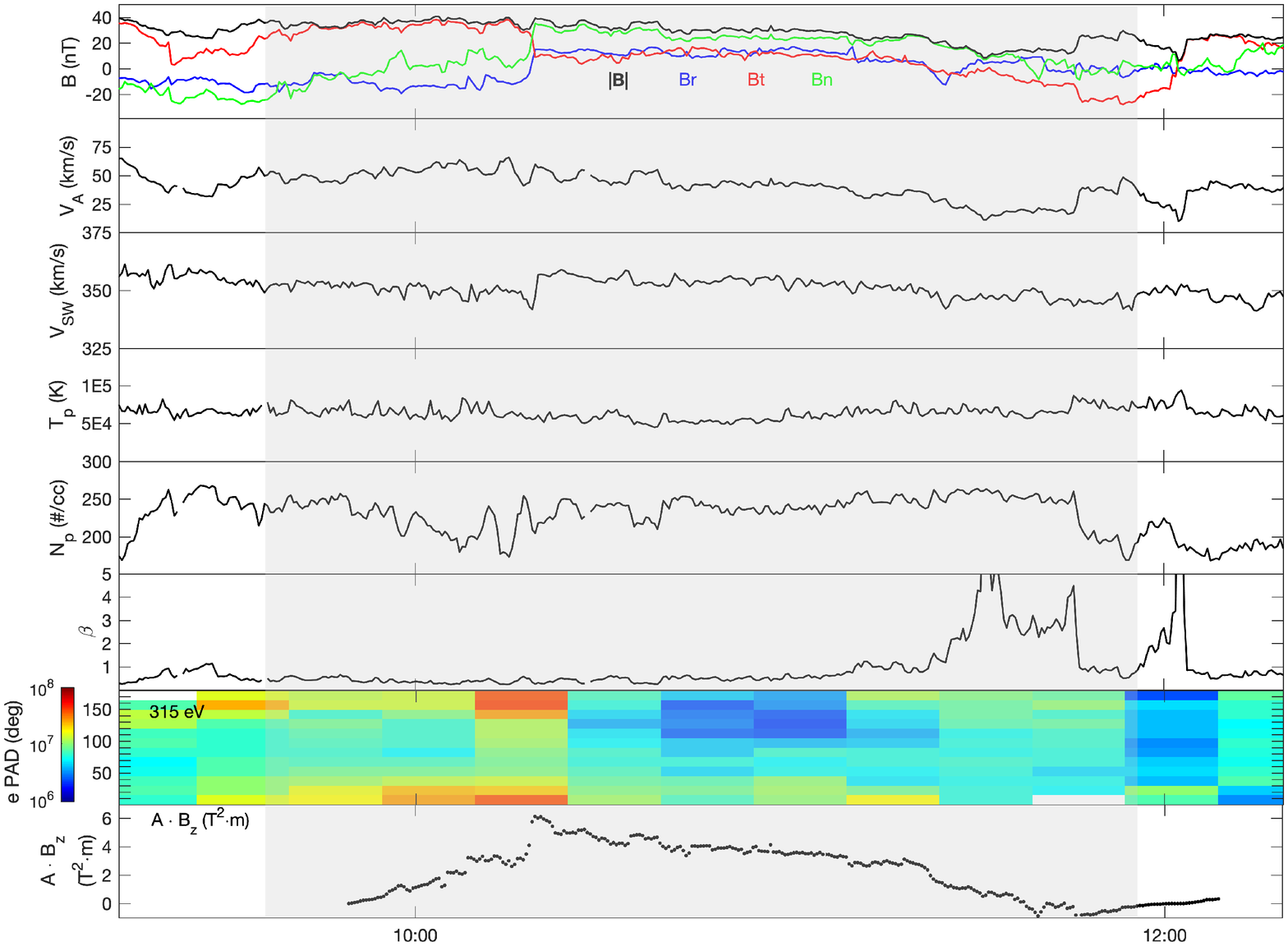}
\includegraphics[width=0.45\textwidth]{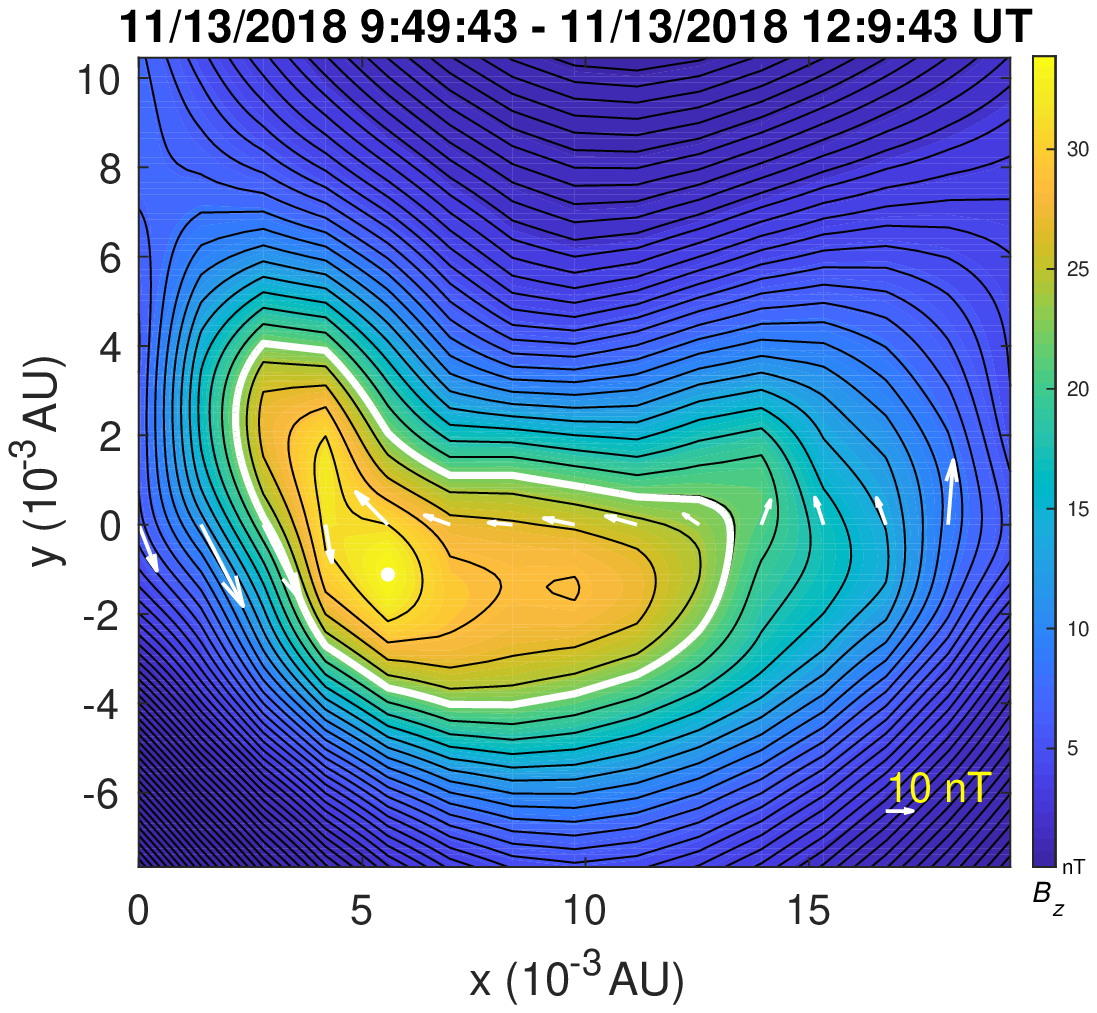}
\includegraphics[width=0.43\textwidth]{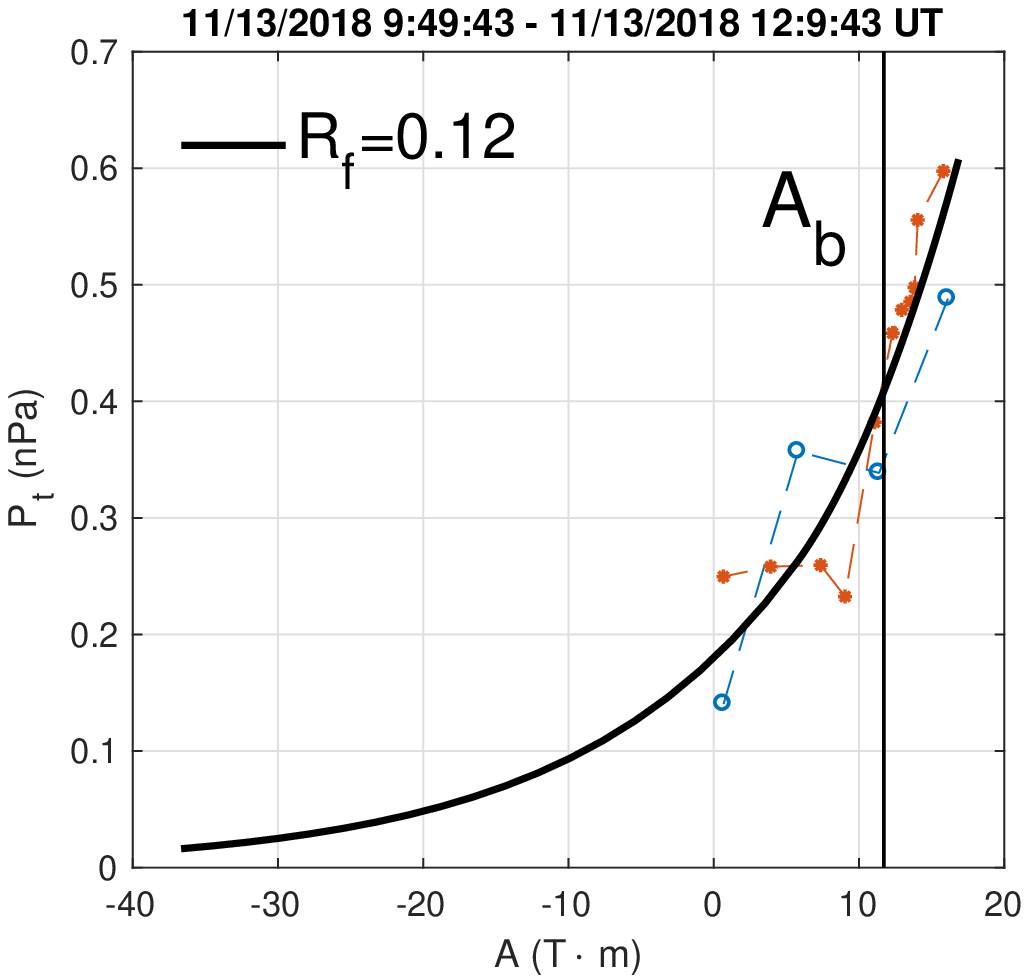}
\caption{Time-series plot and GS reconstruction result of SFR No.~15, 2018 November 13, 9:49:43 - 12:09:43 UT. The flux rope interval is enclosed by the gray block. From the top to the sixth panels: Time-series data; the format follows that of Figure \ref{fig:overview1}. The seventh and eighth panels show the electron PAD and the product of the magnetic flux function $A$ and the axial magnetic field $B_z$, respectively. The bottom left panel is the standard cross-section map from the GS reconstruction with $\hat{\mathbf{z}}=[0.087, 0.492, 0.866]$ in the RTN coordinates. The color represents $B_z$ as indicated by the color bar, while the black contours represent the transverse field lines. The spacecraft path is along the line $y$=0, with the measured transverse field vectors marked by the white arrows. The bottom right panel is the $P_t(A)$ plot: the blue circles, red dots, and the black line represent the measured data points along the spacecraft path from the observations, as well as the fitting curve with the fitting residue $R_f$ as denoted, respectively. The vertical line denoted by $A_b$ marks the magnetic flux function value corresponding to the white contour line in the bottom left panel. }\label{fig:case1}
\end{figure}

\begin{figure}
\centering
\includegraphics[width=0.9\textwidth]{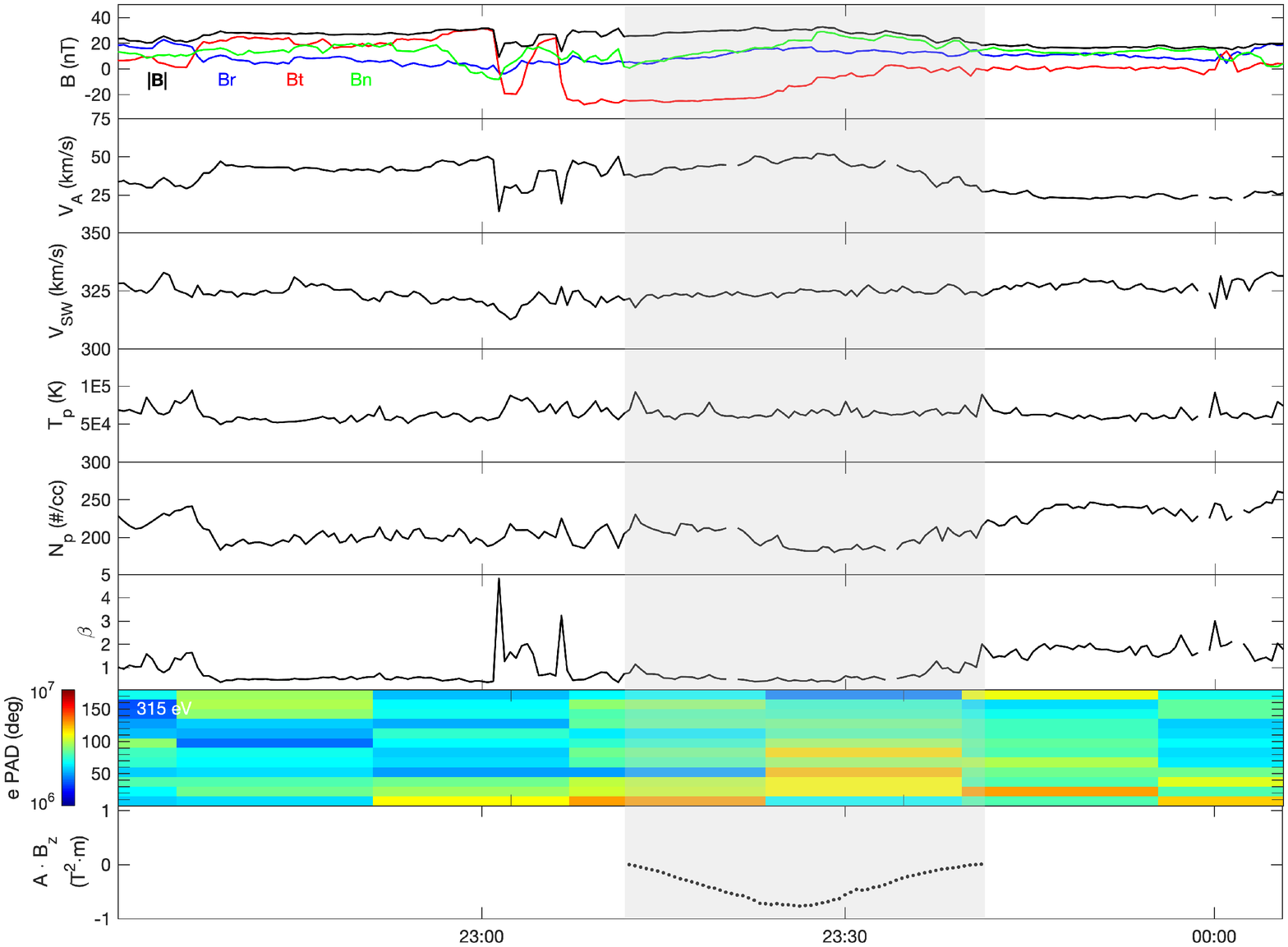}
\includegraphics[width=0.45\textwidth]{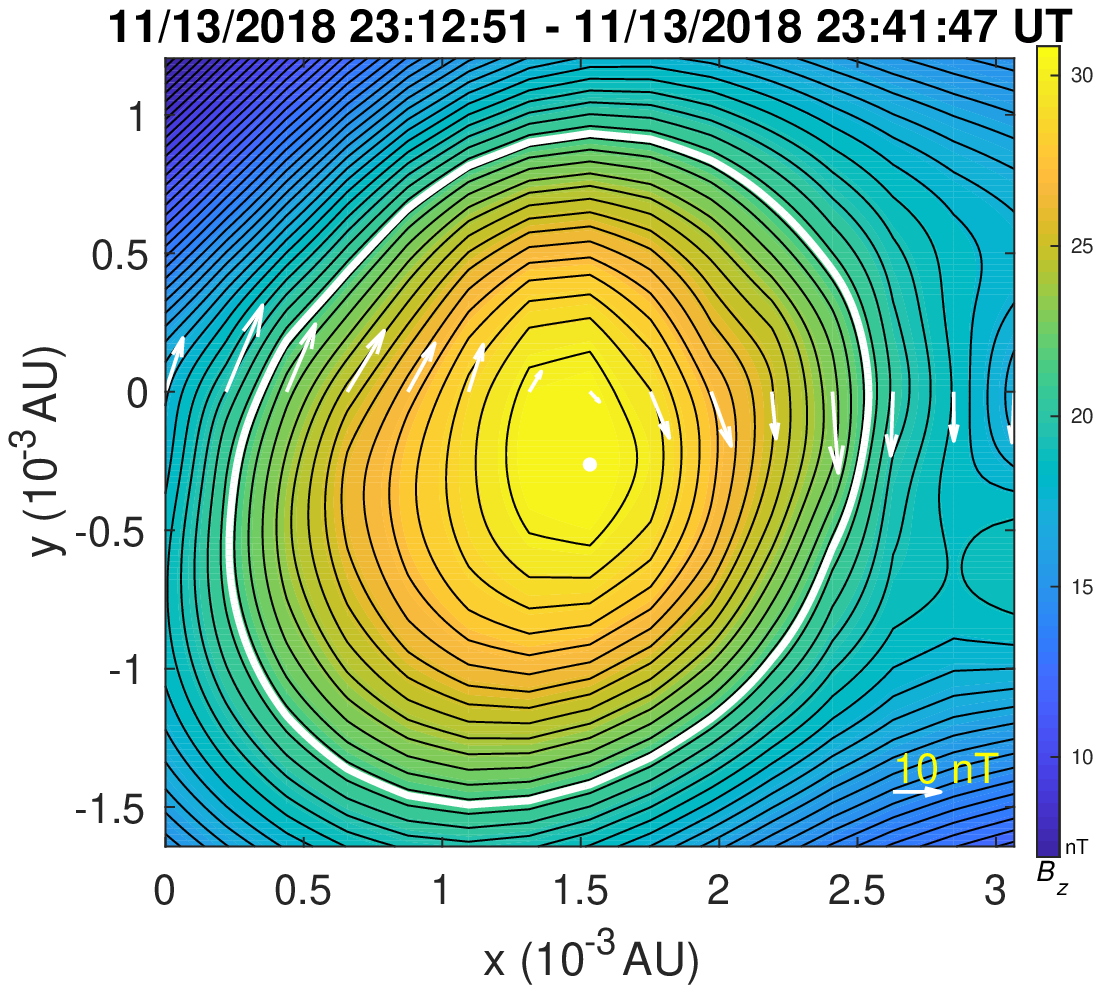}
\includegraphics[width=0.43\textwidth]{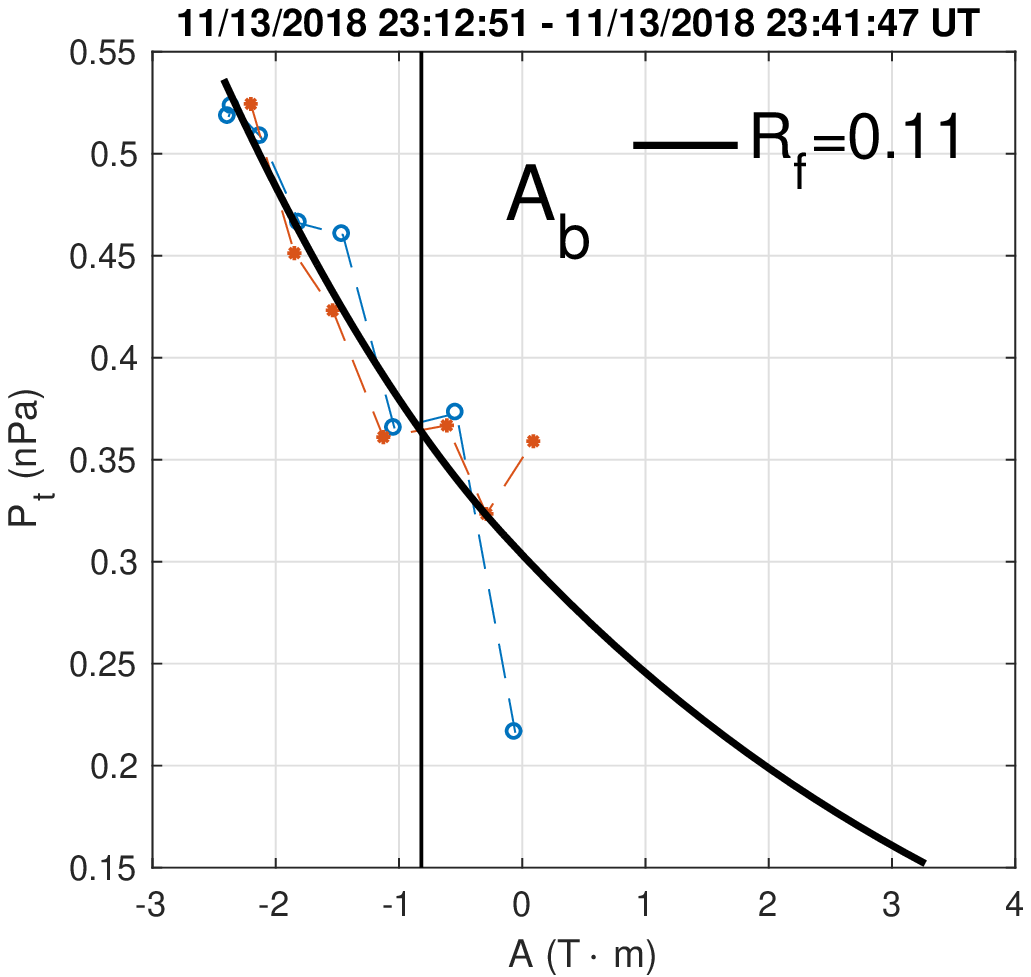}
\caption{Time-series plot and GS reconstruction result of SFR No.16: 2018 November 13, 23:12:51 - 23:41:47 UT, with $\hat{\mathbf{z}}=[0.587, -0.492, 0.643]$ in RTN coordinates. The format follows that of Figure \ref{fig:case1}.}\label{fig:case2}
\end{figure}

Figure \ref{fig:case1} presents the event No.15, which occurred on 2018 November 13. The duration of this SFR is about 140 min, and the scale size is 0.0194 au. The magnetic field components have strong bipolar rotation, while other parameters have merely slight variation except that plasma $\beta$ has significant variation near the end. On the seventh panel, the electron pitch angle distribution (PAD) is plotted. Although the resolution (15 min) is not ideal during this time interval, the electron PAD at 315 eV has relatively strong and intermittent enhancement at both 0$\deg$ and 180$\deg$, some within the identified SFR intervals. The parameter, $A\cdot B_z$, has one positive peak which signifies the time when spacecraft passes the center point of this flux rope. The positive maximum value indicates that the sign of helicity is $+1$, corresponding to right-handed chirality as seen on the cross-section map. Figure \ref{fig:case2} displays another case. In this case, the electron PAD appears to have slight variation only which lacks pronounced features. So do other plasma parameters. The parameter $A\cdot B_z$ has a negative peak and the SFR is with left-handed chirality. These two cases have similar properties as reported in \cite{Zhao2020}. Notice that the magnetic field components in the second case have comparatively weak rotations. This type of SFR can be definitely identified by the GS-based program which is based on more insightful physical considerations than approaches based on visual inspection.

On the other hand, visual inspection is more straightforward in identifying ICMEs and the relatively large-scale flux ropes embedded. A 6-hour ICME (or magnetic cloud) interval at the beginning of 2018 November 12 can be identified from the PSP in-situ measurements \citep[e.g.,][]{Giacalone2020}. \cite{Zhao2020} presented a time-series plot of this structure with the parameters including the magnetic field, the solar wind velocity, the proton number density, and temperature, etc. An ICME flux rope with 264 minutes duration is recognized. On the contrary, \cite{Nieves2020} suggested another scenario that this ICME flux rope probably consists of two flux ropes or a combination of a real and a ``fake" flux rope, instead of being regarded as one large ICME flux rope. Here, by ``fake'' they mean that this structure has similar in-situ signatures to a flux rope but has open field lines.
 
Due to the difference in techniques and criteria, the automatic identification in this study divides the presumably large ICME flux rope interval into three SFRs. Because the detection is carried out by the automated program, which is tailored toward relatively small duration events (usually less than 6 hours; \citet{Hu2018}), the flux rope candidates with the best double-folding patterns are selected, instead of the longest ones which may have relatively “poor” quality as judged by the set of criteria in Table~\ref{table:criteria}. Furthermore, flux rope boundaries also depend on the data and the method of how to process the data. It is known that different methods often yield different boundaries defining a flux rope interval.

\section{Radial Variation of SFRs from 0.16 au to 1 au}\label{sec:radial}

In \cite{Chen2020}, we reported the SFR database via the Helios spacecraft measurements and the associated statistical analysis of SFR properties. The detection was implemented to cover almost the full Helios mission, which lasted from 1975 to 1984 for Helios 1 and 1976 to 1980 for Helios 2. The detection criteria are similar to those listed in Table~\ref{table:criteria} except for the duration range. The Helios time-series data are based on 1 min cadence. Therefore, the duration range starts at 9 min instead of 6 min. The upper limit is also modified to be 2255 min. Although multiple searching windows are applied, most records have duration less than 6 hours, and the mean value is about 25 min. The study of the radial evolution of SFR properties between 0.29 au and $\geqslant$ 1 au becomes feasible and the PSP data further extends the radial distance range below 0.29 au. However given the insufficient number of events, the analysis result presented here should be considered preliminary, which has yet to be improved with additional PSP encounters.

\begin{figure}
\centering
\includegraphics[width=.45\textwidth]{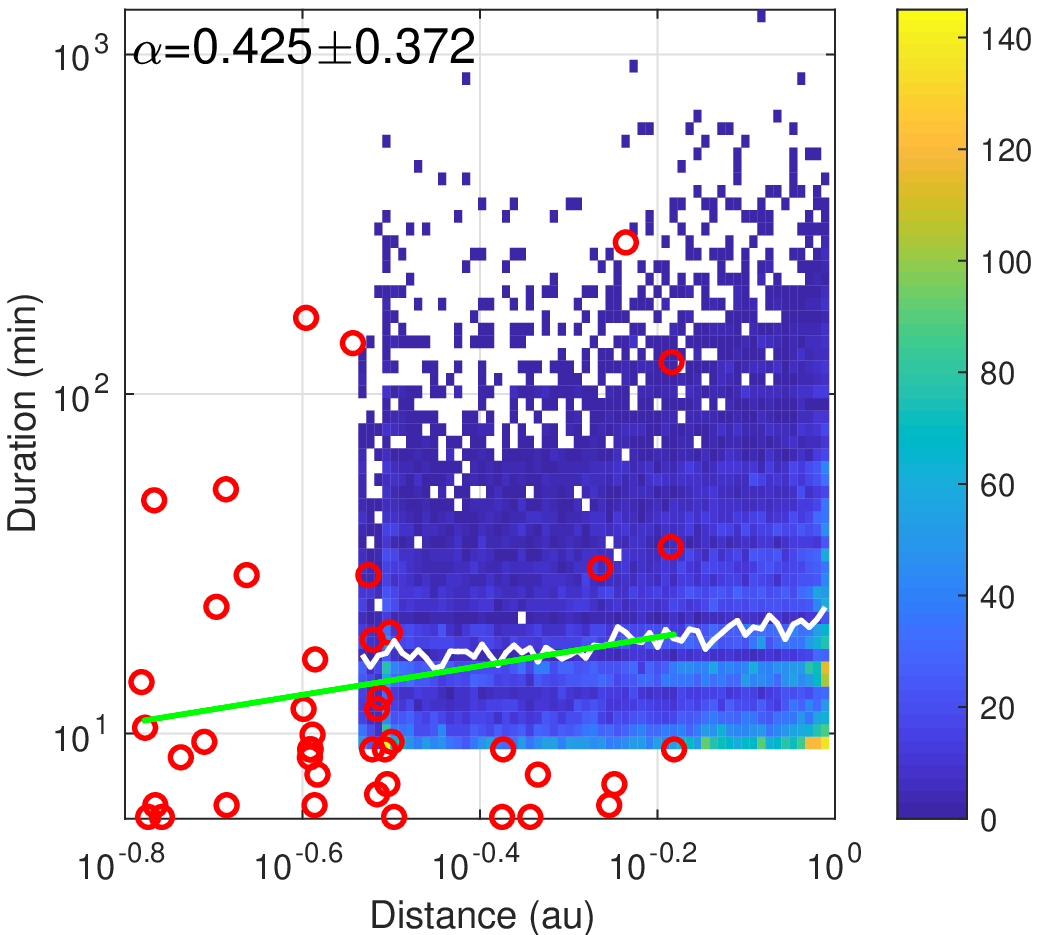}
\includegraphics[width=.45\textwidth]{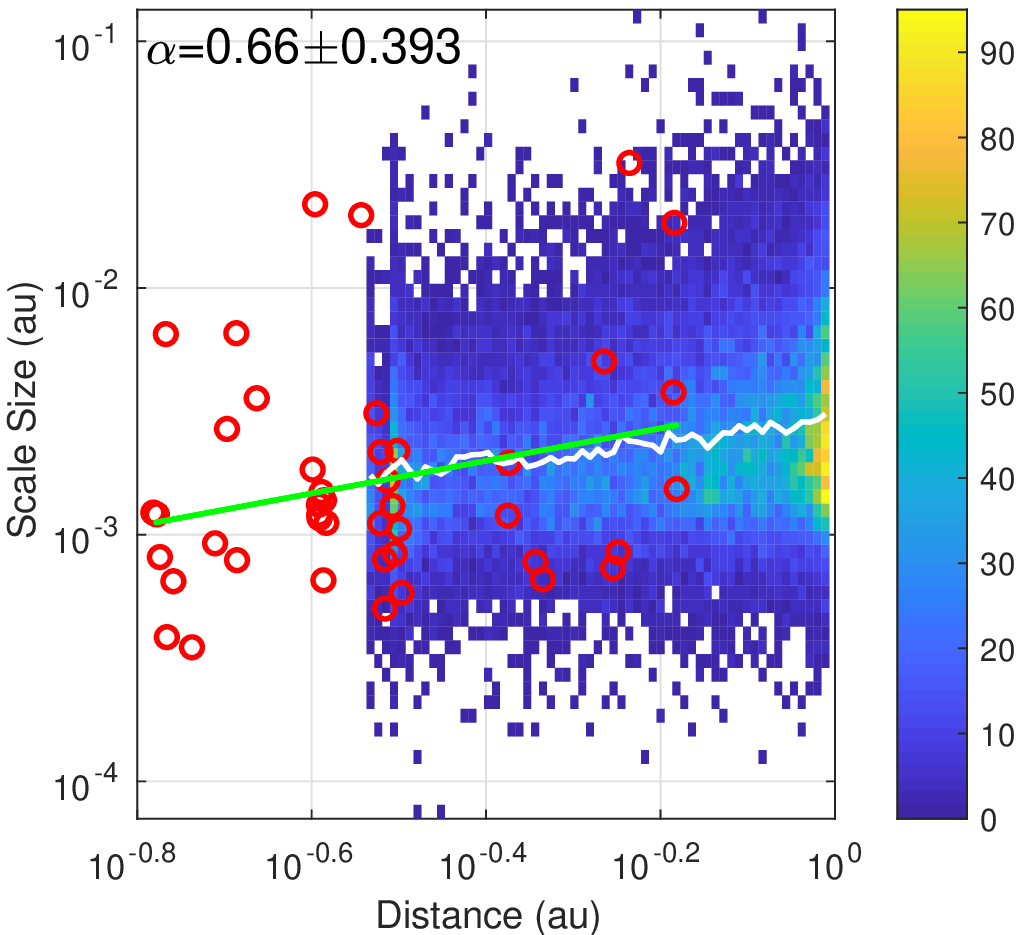}
\caption{Distribution of SFR properties with the radial distances r: (a) duration, and (b) scale size, derived for each SFR. Results via the Helios dataset are presented by the 2D histograms. The bin grids are 60 x 60 in size. All individual records via PSP are directly plotted by red circles. The white curve represents the average value of each bin in $r$, and event count is indicated by the color bar. The green line is a power-law fitting curve for the respective PSP points with the corresponding power-law exponent $\alpha$ denoted on top.}
\label{fig:radial} 
\end{figure}

Figure \ref{fig:radial} exhibits the distributions of SFR parameters, namely, the duration and scale size, with the radial distance $r$ from the Helios SFR database \citep{Chen2020} together with the limited set of PSP results. The bin size in $r$ is set as 0.01 au to account for discontinuous data gaps from 0.29 to 1 au, and the PSP detection result is over-plotted directly, extending down to 0.16 au. As aforesaid, the SFR duration indicates the temporal presence of a structure, whereas the scale size is a measure of the spatial size of the SFR cross-section along the projection of the spacecraft path. In \cite{Chen2020}, we found that both the SFR duration and scale size possess power-law distributions for $r\in$ [0.29, 7-8] au, but with different indices. In Figure \ref{fig:radial}(a) and (b), this conclusion is affirmed by the average value of each bin in $r$ (white curve), showing overall linear variation with increasing $r$ on the log-log scale. Moreover, event counts peak around 0.001 $\sim$ 0.002 au for scale size and 25 min for duration, respectively. 

Although with the first two PSP encounters, additional SFR events can be accounted for smaller $r$, it is far less clear in indicating any discernible trend based on the scattered points in Figure~\ref{fig:radial}. The event count is not sufficient. To guide the eyes, a fitting curve $\propto r^\alpha$ is drawn with the caveat that the standard errors are quite large for $\alpha$. A clear trend of the increase or decrease with the radial distance $r$ for the inner range ($r<0.29$ au) is inconclusive, and has yet to be established by providing more events from upcoming encounters.

\begin{figure}
\centering
\includegraphics[width=.325\textwidth]{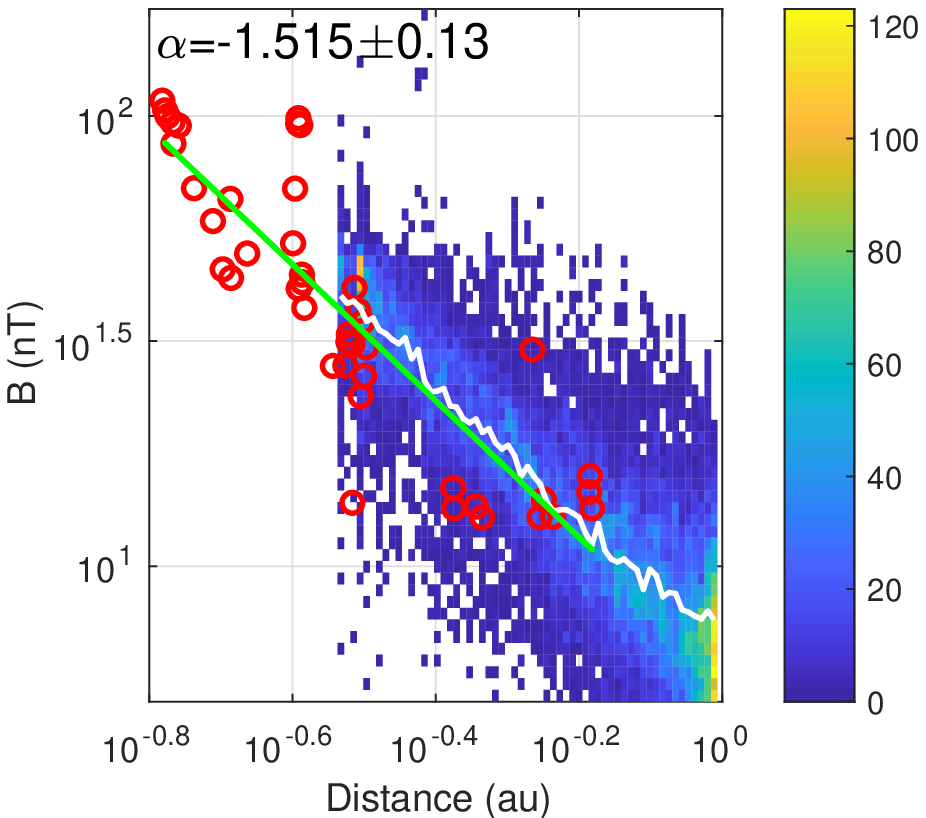}
\includegraphics[width=.325\textwidth]{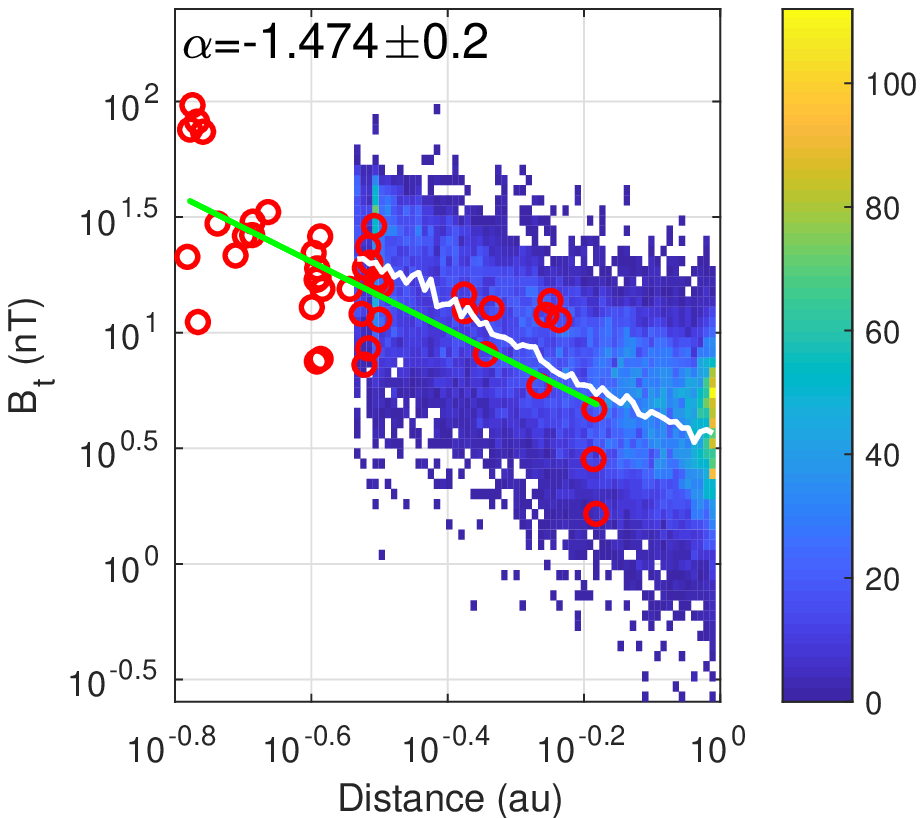}
\includegraphics[width=.325\textwidth]{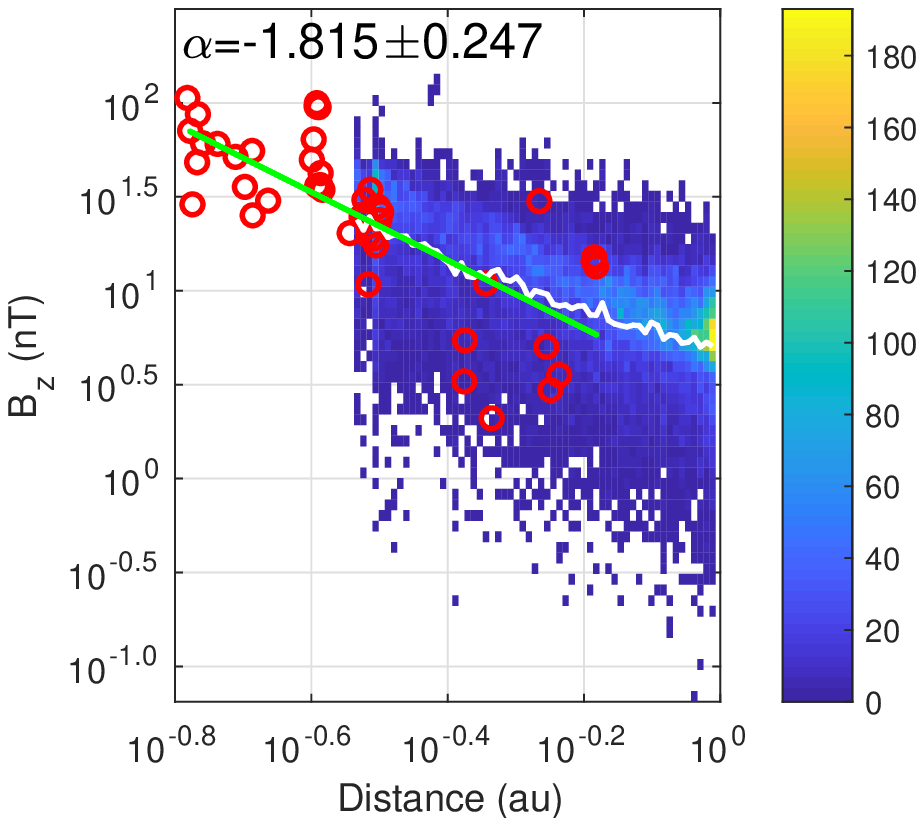}
\caption{Distributions of SFR magnetic field averaged over each SFR interval with the radial distances $r$: (a) the total magnetic field, (b) the transverse field $B_t$, and (c) the axial field $B_z$. The format follows that of Figure \ref{fig:radial}.}
\label{fig:radial1} 
\end{figure}

Figure \ref{fig:radial1} shows the distributions of the various averages of the SFR magnetic field with respect to $r$. The relation between the scattered points and the nominal power-law fitting function (green line) is tighter, as indicated by the fitting results of $\alpha$ with uncertainties. The green line seems to largely follow the decaying trend of the white curve in each panel. The various averages of the SFR magnetic field have evident decaying relations with respect to $r$, and this trend remains valid down to smaller $r$ closer to the Sun. The power-law indices are also approximately the same as the values reported in \cite{Chen2020}, i.e., $\alpha\approx$ -1.4, for $r\ge0.29$ au. 
Although it was speculated that such a consistent and perhaps unified variation seems to comply with the basic background Parker spiral magnetic field \citep{Chen2020}, such a trend may break for the inner range of radial distances, i.e., $r<0.29$ au.

\section{Summary and Discussion}\label{sec:sum}

In summary, we have applied the GS-based automated detection program to the PSP spacecraft measurements and provided the resulting list of SFR records during the first two encounters over the time periods, 2018 October 31 to December 19, and 2019 March 7 to May 15. The new results contain 44 SFRs with duration ranging from 5.6 to 276.3 min. The occurrence rate is compared with 1 au result. An overview of the detection result in the full encounter 1 and part of encounter 2 is presented via time-series plots of measured and derived parameters. With the new event list, some records are discussed further in the context of previous reports on the connection with other structures and cross-check with similar analysis results. Moreover, the SFR database obtained earlier by using the \emph{Helios} spacecraft measurements is cited to investigate the radial variation of SFRs from 0.16 to 1 au combined with the limited number of PSP events from the first two encounters. The main findings are summarized as follows.

\begin{enumerate}
\item Overview of SFRs in the first two encounters reveals that the SFR occurrence rate is far less than that in deep space, owing largely to the prevalence of enhanced Alfv\'enic fluctuations. Such a discrepancy for different heliocentric distances may be an indication of the nature of the local source that generates SFRs through the MHD turbulence. Also, one strict detection criterion, i.e., the Wal\'en test slope threshold, has a significant effect on suppressing the number of events identified.
\item As a consequence, when the PSP traveled closer to the Sun, the occurrence of SFRs becomes scarce to nil, while candidates with high Alfv\'enicity occur more frequently until the PSP left the perihelion. The result in the second encounter shows a little discrepancy from the first encounter, which is probably due to the lack of complete data coverage.
\item A dozen overlapping events with similar boundaries to an alternative list of events are confirmed. The duration distributions for this group of SFRs are similar, and the signs of the magnetic helicity are exactly the same as derived from the two different analysis methods. 
Two examples with opposite signs of helicity out of these well-matched events are presented with their cross-section maps and the corresponding $P_t(A)$ plots. The final GS reconstruction results lend confidence in their 2D cylindrical flux rope configurations.
\item In an early report \citep{Chen2020}, the SFR properties, such as the duration, scale size, and the average magnitude of the magnetic field, have distributions following power laws with different indices. In addition, these properties have clear decaying relations with respect to the increasing radial distance $r$ from 0.29 to 7-8 au. For the limited number of events via the PSP detection, the magnetic field seems to retain these decaying relations. However, the other SFR properties appear to distribute over wider ranges. 
\end{enumerate}

As mentioned earlier, the discrepancy in occurrence rate is owing to the detection criterion of the Wal\'en test slope threshold. Nearly 88\% of candidates are ruled out under a strict limit on the existence of Alfv\'enic structures. Considering that the Alfv\'enicity can change along with the evolution of structure in the solar wind, it is possible for a scenario that Alfv\'enicity reduces when solar wind plasma moves farther away from the Sun \citep{Panasenco2020}. In other words, the event candidates that are recognized with high Alfv\'enicity, i.e., large Wal\'en test slope, close to the Sun may evolve to become quasi-static SFRs at farther distances as Alfv\'enicity decreases.
This scenario also raises uncertainty on the decaying relation between the scale size parameter of quasi-2D non-propagating structures including SFRs with respect to heliocentric distances. The existence of the power-law tendency of the scale size was confirmed mainly for SFRs produced in MHD turbulence in the solar wind over a range of farther distances beyond about 0.3 au. Whether this tendency holds for closer radial distances is still unknown. 

The detection of SFRs in this paper is based on the first two PSP encounters only. The total count of SFRs inevitably affects the current results and is not sufficient to yield statistically significant analysis result, especially for the inner radial distance range ($r<0.29$ au). The future work will be extended to include additional encounters when the PSP mission continues to venture even closer to the Sun. 

\acknowledgments We would like to thank Drs. Anthony Case, Kelly Korreck, and Michael Stevens at CfA for their help with acquiring and processing the PSP data. The PSP data are provided by the NASA CDAWeb. YC and QH acknowledge NASA grants
80NSSC19K0276, 80NSSC18K0622, and NSF grant AGS-1650854 for support. Special thanks also go to the
SCOSTEP/VarSITI program for support of the development
and maintenance of the online small-scale magnetic flux-rope
database website, \url{http://fluxrope.info}.

 \bibliography{bib_database}

\end{document}